\documentclass{aipproc}
\layoutstyle{6x9}

\begin{document}

\title{Options for cosmology at redshifts above one}

\author{Philip D. Mannheim}{address={Department of Physics,
University of Connecticut, Storrs, CT 06269
mannheim@uconnvm.uconn.edu}}

\begin{abstract}

We show that detailed exploration of the $1<z<2$ redshift region can provide
for definitive testing not only of the standard inflationary cosmological
paradigm with its fine-tuned cosmological constant and its mysteriously 
late ($z<1$) onset of cosmic acceleration, but also for the non fine-tuned,
alternate conformal cosmological model, a cosmology which accelerates both
above and below $z=1$. In particular we confront both of these models with
the currently available type Ia supernovae standard candle and  
extended FRII radio source standard yardstick data, with these latter data
being particularly pertinent as they already include a sizeable number of
points in the $1<z<2$ region. We find that both models are able to account
for all available $0<z<2$ data equally well; and with the conformal model
explicitly being able to fit the data while being an accelerating one in 
the $z>1$ region, one is thus currently unable to ascertain whether the
universe is accelerating or decelerating between $z=1$ and $z=2$. To be
able to visualize the supernovae and radio galaxy data simultaneously, we
present a representation of the radio galaxy data in terms of an
equivalent apparent magnitude Hubble diagram. We discuss briefly some
implications of the anisotropies in the cosmic microwave background for
the conformal theory, and show that in that theory fluctuations which set
in at around nucleosynthesis can readily generate the first peak in the
anisotropy data.

\end{abstract}

\maketitle

\section{1.~~Introduction}
\medskip

Through a detailed analysis of type Ia supernovae standard candles
\cite{Riess1998,Perlmutter1999}, of the cosmic microwave background (CMB)
\cite{deBernardis2000,Balbi2000}, and of clusters of galaxies
\cite{Bahcall2000}, standard cosmology has homed in on a rather narrow range
of allowed cosmological parameters, a range centered around
$\Omega_{M}(t_0)=0.3$, $\Omega_{\Lambda}(t_0)=0.7$ or so. While such allowed
values are very encouraging for the standard flat inflationary universe model
\cite{Guth1981}, they are, at the same time, equally deeply troubling for
standard gravity, requiring a fine-tuning of the cosmological constant
$\Lambda$ through  as many as 60 to 120 orders of magnitude, with the fits to
data being altogether disastrous if a value such as $10^{60}$ or
$10^{120}$ for $\Omega_{\Lambda}(t_0)$ were to actually be used. The
currently required value for the cosmological constant associated with an
$\Omega_{\Lambda}(t_0)$ of order one thus poses an extremely severe challenge
to the standard cosmological model which has so far stubbornly resisted
resolution. However, even without any such resolution, it is nonetheless
possible to directly test whether or not nature actually is governed by the
fine-tuned value for $\Lambda$ suggested by the data analysis. Specifically,
since the matter density $\rho_M(t)$ redshifts while the constant $\Lambda$
does not, as one looks back in redshift the relative strength of 
the contributions of these components to the cosmological expansion rate will
vary. In particular, since on its own a normal matter density would lead to
deceleration while by itself $\Lambda$ (if taken to be positive - a further
ad hoc assumption of the standard paradigm) would lead to acceleration, their
inferred current era relative strengths are such that a net cosmic
acceleration is to only be a very late ($z<1$) phenomenon, with the universe
having to be decelerating at all higher redshifts. Study of cosmology above
$z=1$ can thus serve as a major diagnostic for the standard paradigm. And
moreover, as we shall show below, it can also provide for definitive testing
of the fully covariant alternate conformal gravity theory whose cosmology was
originally advocated \cite{Mannheim1990} precisely because it possessed an
underlying symmetry, viz. conformal invariance, which was able to keep the
cosmological constant under control, to thereby lead to a cosmological model
\cite{Mannheim1992,Mannheim1998,Mannheim2000,Mannheim2001a} which was able
to account \cite{Mannheim2001b} for the accelerating universe supernovae
data without any fine-tuning at all while being able to naturally accommodate
a $\Lambda$ as large as elementary particle physics suggests. In this paper
then we therefore explore options for cosmology at redshifts greater than one.

With the acquisition of supernovae data at $z>1$ being quite difficult
(currently there is only one $z>1$ supernova, SN 1997ff, for which both an
apparent magnitude and redshift have been established \cite{Riess2001}),
and with it being some time before the space based SNAP supernovae project will
come on line, it is thus necessary to seek alternate techniques to explore
$z>1$ cosmology. Since data for the very powerful extended FRII radio galaxies
\cite{Fanaroff1974} are already available out to $z=2$ or so (and are readily
extendable to $z=3$), we thus turn to the standard yardstick technique based
on such radio galaxies  which has been developed by Daly and coworkers
\cite{Daly1994,Daly1995,Guerra1998,Guerra2000,Daly2002,Daly2003}, and explore
its implications for cosmology. Even though the technique itself is based on
completely conventional theoretical astrophysical ideas, it is nonetheless
instructive to validate the technique purely by empirical means. Consequently,
we shall first apply the technique to data below $z=1$, and show complete
consistency between its cosmological expectations and those based on the $z<1$
supernovae data themselves. Thus armed, we shall then extend the predictions
of the standard yardstick technique out to $z=2$ and explore its ensuing
implications for cosmology. As such, the procedure that we are following here
is is a well established one in astronomy, namely to check the validity of a
candidate technique against an established one in a given kinematic region,
and to then extend the candidate one into a region which the established one
does not reach. Noting the complementary between the standard yardstick and
standard candle techniques, our analysis thus nicely prepares the $z>1	$
region for its eventual exploration via future supernovae data.

In Sec. (2) we familiarize the reader with the standard yardstick radio galaxy
method, while also presenting a procedure developed jointly with R. A. Daly
which enables us to conveniently represent the standard yardstick technique
predictions in the form of an equivalent apparent magnitude versus redshift
plot, to thus make the method readily visualizable. In Sec. (3) we use the
standard yardstick technique to test some candidate cosmological models, viz.
the standard inflationary ($\Omega_{M}$, $\Omega_{\Lambda}$) model, the
quintessence model \cite{Caldwell1998}, the rolling scalar field model
\cite{Ratra1988}, and the cosmological model based on the alternate conformal
gravitational theory. Through use of such a wide variety of models we are able
to get the broadest possible reading on the interpretation of the data.
Finally, in Sec. (4) we discuss some implications for conformal gravity of
anisotropies in the CMB, another important testing
ground for cosmology, and in work done jointly with K. Horne, show that in the
conformal theory fluctuations which set in at around nucleosynthesis can
readily generate the first peak in the anisotropy data. 

\section{2.~~Theoretical Background}

General use of a standard yardstick for cosmology requires comparing the
measured size of some chosen system at a given redshift with an expected size
for it at that same redshift. However, unlike the common intrinsic
luminosity type Ia supernovae standard candle technique, there does not appear
to be any common intrinsically sized family of astrophysical systems which could
provide a purely empirical analog of the standard candle technique. One thus
has to resort to theory to determine an expected size, with the radio galaxy
method developed by Daly and coworkers relying on standard astrophysical theory
which is independent of cosmology, and with the study of anisotropies in the
CMB using models based on the cosmology itself
to determine the requisite expected size. We discuss the radio galaxy method
here and discuss the CMB technique below.

The primary advantage of the radio galaxy method is that in utilizing the
properties of very powerful FRII classical double radio galaxy sources, one
deals with systems that are luminous enough to permit observation out to $z=2$
and beyond, to thus enable us to go beyond the region currently explored by type
Ia supernovae. The FRII sources consist of an AGN that produces two oppositely
directed supersonically propagating collimated jets which inject energy into
two radio hot spots. The most powerful and least distorted of these radio
sources (referred to as FRIIb by Daly and coworkers) form an unusually
homogeneous population with the average distance between the radio hotspots,
$<D>$, at a given redshift exhibiting a rather small dispersion. $<D>$ then
serves as a requisite measured size, and with these sources subtending a small
opening angle $\theta$ at the observer, we can set $<D>=\theta R(t)r=\theta
R(t_0)r/(1+z)$ where $r$ is defined by the Robertson-Walker null
geodesic relation
$\int_{t}^{t_0}cdt/R(t)=\int_0^{r}dr/(1-kr^2)^{1/2}$. 

On assuming that the supersonic flow (of average rate of growth $v_L$) of the
FRIIb sources can be described by strong shock physics, and assuming that the
total lifetime of the source is related to the beam power, $L$, of the source
according to the power law $t_*\propto L^{-\beta/3}$, Daly and coworkers show
(see e.g. \cite{Daly2003}) that the size $D_*=v_Lt_*$ to which such an FRIIb
system will eventually be expected to grow is related to $<D>$ as
\begin{equation}
<D>/D_*=k_0y(z)^{(6\beta-1)/7}[k_1y(z)^{-4/7}+k_2]^{\beta/3-1}
\label{1a}
\end{equation}
where $y(z)=H_0R(t_0)r/c$, and where $k_0$, $k_1$ and $k_2$ are specific
(though rather complicated) functions of observables associated with the FRIIb
systems which are given in \cite{Guerra2000}. In order to apply Eq. (\ref{1a})
to cosmology Daly and coworkers assume further that the expected $D_*$ will
be universally proportional to the measured average size $<D>$ of all of the
sources in the parent sample which are at the same redshift as the given
source, i.e. that the ratio $<D>/D_*$ is a redshift independent constant
$\kappa$.\footnote{The absolute normalization of $\kappa$ is unimportant.
What matters for the technique is that $\kappa$ be independent of
redshift, something which is borne out in the fits.} At the present time a
parent population of 70 FRIIb radio galaxies has been identified, with 20 of
the sources having been observed in detail. The full 70 source parent sample is
thus used to determine $<D>$, while values for $D_*$ are determined from the
well studied 20 sources. Even though none of the assumptions which go into this
analysis is particularly contentious or unreasonable, nonetheless regardless of
the validity of its theoretical underpinnings, the radio galaxy standard
yardstick technique can be considered independently verified by the successful
comparison between the $z<1$ radio galaxy and supernovae data given below, a
comparison which provides fitted values for the phenomenological $\beta$ and
$\kappa$.

While the fits to be given in the tables below are based on the use of the full
theoretical calculation outlined above, should energy losses due to
inverse Compton cooling of relativistic electrons by CMB photons be negligible
(something thought to be a good though not perfect approximation for the
available $z<2$ FRIIb sample), the parameter $k_2$ in Eq. (\ref{1a}) can then
be neglected, with Eq. (\ref{1a}) then simplifying to
\cite{Guerra1998} 
\begin{equation}
\kappa=(R(t_0)r)^{g(\beta)}Q
\label{3a}
\end{equation}
where $g(\beta)=3/7+2\beta/3$ and where
$Q=(H_0/c)^{g(\beta)}k_0k_1^{\beta/3-1}$  depends only on observed quantities.
If we now define a quantity
\begin{equation}
m_{RG}=5 {\rm log}[(1+z)Q^{-1/g(\beta)}]+(5/g(\beta)){\rm
log}~\kappa+M+5{\rm
log}[H_0/c]~~,
\label{4a}
\end{equation}
where $M=M_B-5{\rm log}(H_0/c)+25$, we can then introduce a convenient
equivalent or effective apparent magnitude parameter $m_{RG}$ for radio
galaxies, viz.
\begin{equation}
m_{RG}=M+5 {\rm log}[(1+z)~H_0R(t_0)r/c]~~,
\label{5a}
\end{equation}
an expression completely analogous to the B band Hubble diagram relation
\begin{equation}
m_{B}=M+5 {\rm log}[(1+z)H_0R(t_0)r/c]~~~.
\label{6a}
\end{equation}
The parameter $M$ introduced here is related to the intrinsic absolute
magnitude $M_B$ of type Ia supernovae via the relation $M_B= M -40 -5 ~{\rm
log}(3h^{-1})$, where $h$ is the Hubble constant as measured in units of 100
km/s/Mpc. For $h = 0.65$, $M_B = M - 43.32$, the typical measured value of $M
=23.95$ given below corresponds to the value $M_B = -19.37$ found in the
supernovae data analyses themselves \cite{Riess1998,Perlmutter1999}. By
putting the radio galaxy data into the same format as the supernovae data we
can thus plot both data sets on one and the same graph, and while this is
only an approximate procedure, it nonetheless permits an easy visualization
of the entire $0<z<2$ region.\footnote{Recently Daly \cite{Daly2003} has
generalized this formalism by using the full Eq. (\ref{1a}) to extract the
dependence of $y(z)$ on $z$ directly without needing to make any
approximation at all. Since $y(z)=(H_0/c)d_L/(1+z)$  where $d_L$ is the
luminosity distance, a plot of $y(z)$ against $z$ is essentially a Hubble
plot, with the plots given in \cite{Daly2003} being found to
exhibit the same general trends as those given in Figs. 1 and 2 below.} 

Application of the theory to data requires the specification of a global 
cosmological model based on a theory of gravity. For standard gravity with
a set of perfect fluid sources each with an equation of state
$p_i(t)=w_i\rho_i(t)$ and with an $\Omega_i(t)$ parameter
given by $\Omega_i(t)=8\pi G\rho_i(t)/3c^2H^2(t)$, the coordinate
distance is given in the canonical $k=0$ universe case by the familiar
\begin{equation}
R(t_0)r={c \over H_0} \int_0^z {dz \over [\sum_i
\Omega_i(1+z)^{3+3w_i}]^{1/2}}~~, ~~\sum_i \Omega_i=1~~.
\label{7a}
\end{equation}
Eq. (\ref{7a}) encompasses not only a standard inflationary universe
with a $w_M=0$, $\Omega_M>0$ matter fluid and a $w_{\Lambda}=-1$,
$\Omega_{\Lambda}>0$ cosmological constant, but also a quintessence model with
a matter fluid and an $\Omega_Q>0$ quintessence fluid whose $w_Q$ is negative.
With slight adjustment Eq. (\ref{7a}) can also be applied to the rolling scalar
field model \cite{Ratra1988}, with its power law potential
$V(\phi)\propto \phi^{\alpha}$ leading to a $w$ parameter which then depends
on $z$.

As well as study the standard theory, we shall also explore the fully
covariant alternate conformal gravitational theory, a theory which sets out
to solve some of the most troubling problems in astrophysics, viz. the
cosmological constant and dark matter problems, by modifying gravity rather
than by making ad hoc adjustments to the energy-momentum tensor. While the
conformal theory is found to recover the results of standard gravity for
solar system sized distances or less, its departure from the standard theory
on larger distance scales has enabled it to naturally resolve the dark matter
and dark energy problems without any fine-tuning at all,
with its only known difficulty (a point we return to below)
being an inability to nucleosynthesize sufficient primordial deuterium. 

In the conformal theory it is found that even while the low energy limit of
the theory is controlled by a dynamically induced but otherwise standard
attractive Newton constant $G$, its cosmology is controlled by an entirely
different induced gravitational constant $G_{eff}$, a repulsive rather
than attractive coupling constant which is  given as $G_{eff}=-3c^3/4\pi \hbar
S_0^2$ where $S_0$ is the (very large) expectation value of a scalar urfield
which is to spontaneously break the conformal symmetry cosmologically. Apart
from this specific change the cosmological evolution equation is otherwise
completely standard, taking (for a matter density which redshifts as
$1/R^n(t)$) the form
\cite{Mannheim1992,Mannheim1998,Mannheim2000,Mannheim2001a}
\begin{equation}
\dot{R}^2(t)+kc^2=\dot{R}^2(t)[\bar{\Omega}_M(t)+\bar{\Omega}_{\Lambda}(t)]
~,~~q(t)=(n/2-1)\bar{\Omega}_M(t)-\bar{\Omega}_{\Lambda}(t)
\label{8a}
\end{equation}
with $\Omega_i(t)$ having been replaced by $\bar{\Omega}_i(t) =8\pi
G_{eff}\rho_i(t)/3c^2H^2(t)$. Moreover, in the conformal theory the sign of
the parameter $\Lambda$ is explicitly known to necessarily be negative
since it arises from elementary particle physics phase transitions which
occur as the universe cools down, i.e. by transition from an unbroken
symmetry phase with $\Lambda=0$ to a broken one with a lower energy. Then, with
$G_{eff}$ also being  negative, it follows that $\bar{\Omega}_{\Lambda}(t)$
itself must necessarily be positive. Moreover, with $\bar{\Omega}_{M}(t)$
necessarily being negative precisely because $G_{eff}$ is negative, it
follows that $q(t)$ is then always negative, with conformal cosmology thus
automatically being an accelerating one in each and every epoch no matter how
big or small $\Lambda$ might be. As regards the numerical value of
$\bar{\Omega}_{\Lambda}(t)$, we note further that the larger $S_0$ the smaller
$G_{eff}$, and thus the smaller the amount by which the cosmological constant
gravitates. Thus in the conformal theory it is not the cosmological constant
which gets quenched but rather its effect on cosmic evolution, with the amount
of gravity produced by a matter source being radically reduced from the
amount generated by the same source in the standard theory. Moreover, this
quenching is done by the theory itself without the need for any fine-tuning,
leading to a theory in which no matter how huge $\Lambda$ might be,
$\bar{\Omega}_{\Lambda}(t)$ always \cite{Mannheim2000,Mannheim2001a} has to
lie between zero and one in all epochs except the very earliest (the only
epoch where $\bar{\Omega}_M(t)$ is of consequence even though $\rho_m(t)$
contains the completely standard amount of luminous material), with the late
universe deceleration parameter being given as
$q(t)=-\bar{\Omega}_{\Lambda}(t)$, so that at late times $q(t)$ then has to
automatically lie between zero and minus one no matter what. The theory thus
gives a controlled amount of cosmic acceleration in each and every late
universe epoch without any fine-tuning at all, with it being the absence of any
decelerating epoch above
$z=1$ which serves as a clear discriminator between it and the standard
theory. Given Eq. (\ref{8a}), the conformal
theory coordinate and luminosity distances are then found to be given by
\cite{Mannheim2001a,Mannheim2001b}
\begin{equation}
R(t_0)r={d_L \over (1+z)}=-{c(1+z) \over H_0q_0}\left[ 1- \left\{1+q_0- {q_0
\over (1+z)^2}\right\}^{1/2}\right]~~,
\label{9a}
\end{equation}
to thus give a one parameter family of fits labelled by the current value
of $q_0$, a value which, as we just noted, has to necessarily lie between zero
and minus one. Given Eqs. (\ref{7a}) and (\ref{9a}) we turn now to the
data.

\section{3.~~Options for cosmology at redshifts above one}
\medskip

For the supernovae data we follow the authors \cite{Perlmutter1999} and fit 38
of their 42 reported data points together with 16 of the 18 earlier lower $z$
points of \cite{Hamuy1996}, for a total sample of 54 $z<1$ supernovae data
points. (While we thus leave out 6 questionable supernovae data points for the
fitting, nonetheless, for completeness we still include them in the displayed
Fig. 1.) For the radio galaxies we use the 14 data points listed in
\cite{Guerra1998} and the 6 listed in \cite{Guerra2000}, for a total of 20
radio sources. Of the models for which we provide fits below, the implications
for the radio galaxy data of three of them have already been well studied in the
literature by Daly and coworkers, with the standard model radio
galaxy predictions having been given in \cite{Guerra2000}, the quintessence
model predictions in \cite{Daly2002} and the scalar field model predictions
in \cite{Podariu2002}. Fits to the supernovae data using the standard model and
the quintessence model abound in the literature, with supernovae fits
using the scalar field model having been given in \cite{Podariu2000}, and
supernovae fits using conformal cosmology having been given in
\cite{Mannheim2001b}. As far as all of those published fits are concerned what
is new here is only in the way the fits are organized in the tables below,
excepting that the conformal cosmology radio galaxy fits are new.\footnote{The
fits themselves were prepared for the author by R. A. Daly and M. P. Mory
using Daly's master program which can generate radio galaxy fits for 
assigned cosmologies, and the author is altogether indebted to them for
doing so.}

For the $k=0$ standard model fitting to the supernovae data we recover the
results of \cite{Perlmutter1999}, and obtain a minimum $\chi^2=56.76$ with
$\Omega_M=0.3$, $\Omega_{\Lambda}=0.7$, with the 68\% confidence region being
given as $0.2<\Omega_M<0.35$, $0.8>\Omega_{\Lambda}>0.65$ (see Table 1, where
our identification of 51 rather than 52 degrees of freedom is due to an internal
aspect of the data extraction procedure used in \cite{Perlmutter1999}). Of the
20 radio sources 9 have $z<0.9$, and a fit to them alone yields a best fit
$\chi^2=11.65$. Combining now the two $z<1$ data sets then yields a best fit
$\chi^2=68.72$ for the 63 points with $z<1$, where now the minimum is at
$\Omega_M=0.25$, $\Omega_{\Lambda}=0.75$, with the 68\% confidence region being
given as $0.2<\Omega_M<0.35$, $0.8>\Omega_{\Lambda}>0.65$. Noting the complete
overlap of the fitting parameters and noting that $56.76+11.65=68.41$ is
extremely close to $68.72$, we thus find complete compatibility between the
cosmologies implied by the $z<1$ supernovae and  radio galaxy data. And with
this very same concordance being found in the $z<1$ analyses of all of the
other cosmological models being considered here, we believe that one may
therefore regard the standard yardstick technique as having been empirically
confirmed.

Having established the credentials of the radio source technique, we now
include the 11 radio galaxies with $z>1$ and make an overall standard model fit
to all 74 of the $z<2$ data points. We find a best fit $\chi^2=74.41$ for the 74
points with $z<2$, where now the minimum is at $\Omega_M=0.25$,
$\Omega_{\Lambda}=0.75$, with the 68\% confidence region being given as
$0.2<\Omega_M<0.35$, $0.8>\Omega_{\Lambda}>0.65$. With a best fit $\chi^2=16.89$
being found for the 20 radio galaxy data points alone and with
$56.76+16.89=73.65$ being within one standard deviation of $74.41$, we again
find complete compatibility between the standard yardstick and standard candle
approaches. 

In an examination of the fits it was found that a huge amount of the $\chi^2$
was contributed by just one radio galaxy, viz. 3C 427.1 at $z=.572$, an outlier
which is more than 3 $\sigma$ away from the best fits.  Consequently, we also
investigated fits to the data with this potentially questionable source removed,
with the resulting outcome for the standard inflationary cosmology being listed
in Table 5 and displayed in Fig. 1 as an equivalent apparent magnitude fit
and then in Fig. 2 as a residual equivalent apparent magnitude fit
with respect to the convenient empty universe  baseline.\footnote{While not
included in the fit, for completeness we have still included 3C 427.1 in the
figures. The horizontal error bars which are shown in the fits to the radio
galaxy data points are obtained by combining the uncertainties in the data and
the mean fitted deviations of the parameters $\beta$, $\kappa$ and $M$ as given
in column 5 of Table 5. (Details of this error bar analysis are given in
\cite{Podariu2002}.)} With the plot of the altogether acceptable fitting of Eqs.
(\ref{4a}) - (\ref{6a}) to the data being shown in Fig. 1, we believe that this
figure can reasonably be interpreted as an early look at the
$z<2$ Hubble diagram.

For comparison purposes we have also included in Fig. 1  a plot of the 
apparent magnitude expectations associated with the illustrative ($\Omega_M=1$,
$\Omega_{\Lambda}=0$, $\Omega_{k}=0$), ($\Omega_M=0$, $\Omega_{\Lambda}=0$,
$\Omega_{k}=1$), and ($\Omega_M=0$, $\Omega_{\Lambda}=1$, $\Omega_{k}=0$) models
(as calculated from Eq. (\ref{6a}) with $M=23.95$). The curvature dominated
($\Omega_M=0$, $\Omega_{\Lambda}=0$, $\Omega_{k}=1$) empty universe is a
coasting one with $q(t)$ being zero in all epochs, and is thus a particularly
convenient  baseline, a point we emphasize in the residual magnitude plot with
respect to it given in Fig. 2. As we see, the ($\Omega_M=0.25$,
$\Omega_{\Lambda}=0.75$) model apparent magnitude crosses this baseline at
around $z=1.6$,\footnote{Because the dependencies on redshift of the
deceleration parameter and the luminosity distance are quite different, for 
given assigned values of $\Omega_M(t_0)$ and
$\Omega_{\Lambda}(t_0)$ the parameter $q(z)$ can change sign at a much lower
redshift than the one at which $d_L(z)$ would cross the empty universe baseline.}
a somewhat higher (rather than lower) value than the $z=1.3$ value where an
($\Omega_M=0.3$, $\Omega_{\Lambda}=0.7$) universe would cross, to thus indicate
that no compelling case from the radio galaxy data can be made that the $z>1$
region is any less cosmically repulsive than the $z<1$ region. Additionally in
the figures we have included the SN 1997ff data point at
$z=1.7\pm^{0.10}_{0.15}$, and we see that the radio sources (and particularly the
radio sources at the highest available redshifts) are not supporting its
suggestion that cosmic repulsion is in fact weakening above $z=1$. As regards
SN 1997ff, we additionally recall that the authors of \cite{Riess2001} had noted
that this particular supernova just happened to be lensed by two foreground
galaxies along the line of sight, so it might well be a lot dimmer than
indicated in the figures, an effect which would then move it more toward the
cosmically repulsive side of the empty universe baseline. From Fig. 2 we
additionally infer that by extending the Hubble diagram out just a little bit
further in $z$, it should then rapidly become apparent whether the
standard model $z=1.6$ crossover is supported by higher $z$ data. Moreover, even
without this, filling in the $z<2$ region with more data points could itself
already sharply constrain cosmology.

We have also made fits to the data using a quintessence model, a scalar field
model and a conformal cosmology model, and display their best
fits in Tables 2, 3 and 4 (3C 427.1 included) and Tables 6, 7 and 8 (3C 427.1
excluded), and plot their best outlier excluded fits in Figs. 1 and
2.\footnote{For fitting reasons we have constrained the quintessence $w$
parameter to the range $-3\leq w\leq 0$, and the scalar potential parameter
$\alpha$ to the range $0 \leq\alpha \leq 8$.} As we see, both the quintessence
model fitting and the scalar field model fitting are every bit as acceptable as
the standard ($\Omega_M$, $\Omega_{\Lambda}$) model fitting, with their best
$z>1$ fits also not being found to be any less cosmically repulsive than those
for $z<1$ (if anything both of the models go in the direction of making
$\Omega_M$ smaller). This suggestion of a potentially continuing cosmic
repulsion above $z=1$ is also shared by the conformal gravity fits, fits which
are just as good as the standard model and quintessence fits while being
strictly on the repulsive side of the empty universe baseline at all $z$. The
conformal gravity fitting to the 74 total data points which we present here is
completely consistent not only with the earlier conformal gravity fitting to the
54 $z<1$ supernovae data points given in \cite{Mannheim2001b}, but also with the
$z>1$ predictions made in the same paper; with the very success of the
conformal gravity fitting that we have presented here implying that it is not
yet possible to ascertain whether the universe is actually accelerating or
decelerating between $z=1$ and $z=2$,\footnote{A recently updated analysis of
the lensing of SN 1997ff now indicates \cite{Benitez2002} that SN 1997ff was
probably magnified even more than had previously been thought, to thus
necessitate repositioning it even further toward the cosmically repulsive side
of Fig. 2; with the authors of \cite{Benitez2002} noting that the conformal
cosmology prediction for this supernova would then be brought well within the 2
$\sigma$ level of acceptability.} thus making any extension or filling in of
the Hubble diagram potentially highly instructive. The conformal gravity fits
are also significant in that at the present time they are in fact the only non
fine-tuned fits to the accelerating universe data that have so far been
presented in the literature, to thus at the very least show that it is in
principle possible to fit the data without fine-tuning, with the currently
available $0<z<2$ data not at all rejecting the only predictive cosmological
model presented so far in the literature in which the cosmological constant
problem is naturally solved. 

\section{4.~~Options for cosmology at recombination}
\medskip

Other than the $0<z<2$ region, the two other primary regions where cosmology can
be tested are the nucleosynthesis era and the CMB
recombination era, studies of which lead in the standard theory to remarkably
successful fitting associated with an $\Omega_M(t_0)=0.3$,
$\Omega_{\Lambda}(t_0)=0.7$ universe. As we had indicated earlier such a
universe can directly be tested in the $1<z<2$ region. However, because it might
be some time yet before a definitive answer to such testing is actually
obtained, and because the cosmological constant problem associated with such an
$\Omega_M(t_0)=0.3$, $\Omega_{\Lambda}(t_0)=0.7$ universe is so very severe, it
is of value to ask whether the CMB anisotropy data could admit of any alternate
explanation. As well as a being posed simply as a general question (namely, how
much of the success of the CMB fitting is due to detailed features of a
particular model and how much might be generic), one can also ask how well any
candidate alternate theory might fare. While the cosmological fluctuation theory
associated with the alternate conformal gravity theory being considered in this
paper has yet to be fully developed, a first step in this regard has recently
been taken by the author and K. Horne, one we now report on.

Basic to the CMB analysis is a determination of the true proper diameter
$d(\theta)$ of some candidate yardstick at coordinate $r$ and redshift
$z=R(t_0)/R(t)-1=T(t)/T(t_0)-1$ which subtends an angle $\theta$ at an observer
at $r=0$, $z=0$, a proper diameter which for a general Robertson-Walker geometry
is given by
\begin{equation}
d(\theta)=2R(t)\int _0^{r{\rm sin}(\theta/2)}{dr \over
(1-kr^2)^{1/2}}~~,
\label{10a}
\end{equation}
and which for small $\theta$ reduces to the relation $d(\theta)=\theta R(t)r$
used earlier for the radio galaxies. If the yardstick used for the CMB
analysis is due to the growth of some cosmological fluctuation which started at
some earlier fluctuation time $t_{F}$, the proper distance
$D(t,t_F)$ of the fluctuation at the time $t$ will be given by
\begin{equation}
D(t,t_F)=R(t) \int _{t_{F}}^{t} {dt \over R(t)}~~, 
\label{12a}
\end{equation}
so that a comparison of the measured $d(\theta)$ with a model choice for
$D(t,t_F)$ allows one to test and constrain the chosen model.

Since the treatment of the isotropy of the CMB in the conformal theory differs
substantially from the discussion in the standard theory (the conformal
cosmology CMB derived from Eq. (\ref{8a}) is already causally connected
\cite{Mannheim1998} even without any inflationary phase), it is instructive to
first recall the standard model discussion. With the largest possible value for
$D(t,t_F)$ being given by a fluctuation which set out at $t_{F}=0$, for a
standard $k=0$ cosmology which is radiation dominated ($R(t)=At^{1/2}$) until
recombination, at recombination the maximum $D(t_R,t_F)$ is then given as
$D(t_R,t_F=0)=2t_R$. Similarly, for a $k=0$ standard cosmology which is matter
dominated ($R(t)=Bt^{2/3}$) since recombination, $d(\pi)$ is given as
$d(\pi)=6t_R^{2/3}t_0^{1/3}$, with the ratio
$D(t_R,t_F=0)/d(\pi)=(T(t_0)/T_R)^{1/2}/3$ thus being very much less than one.
Thus despite the high isotropy found for the CMB, in a $k=0$ Robertson-Walker
universe opposite points on the CMB sky would not be causally connected, with
the angle $\theta=2(T(t_0)/T_R)^{1/2}/3$ subtended on the sky by $D(t_R,t_F=0)$
being of order only $1^{\circ}$. In the standard theory this causality problem
is solved by having an inflationary de Sitter phase occur very early in the
history of the universe prior to the onset of the Robertson-Walker phase. This
inflationary phase not only reconciles the causality conflict, it also
converts what was a considerable difficulty into a potentially considerable
triumph, since the very same $D(t_R,t_F=0)$ then no longer sets the scale for
the isotropy of the CMB, but rather for its anisotropy as caused by fluctuations
generated during the very same inflationary era. With the detection of an
anisotropic peak in the CMB associated with precisely such a $1^{\circ}$ scale,
it is now taken as a given that the fluctuations which are seen in the CMB must
indeed have originated in the very early universe, with the very success of
inflation in describing the CMB leading one to conclude that the spatial
3-curvature of the universe is having a negligible effect on current era cosmic
expansion so that $\Omega_k(t_0)=-kc^2/\dot{R}^2(t_0)$ is
negligible.\footnote{In passing we note that inflation only quenches
$\Omega_k(t)$, but does not actually fix a value for $k$ itself. What is to then
fix the global topology of the universe in the standard theory is yet to be
identified.} However, as we shall now show, in theories such as conformal
cosmology whose Robertson-Walker phase already is causal, an altogether
different option is possible. 

With only the conformal matter density $\bar{\Omega}_{M}(t)$ contributing to the
conformal evolution equation Eq. (\ref{8a}) at temperatures above the
phase transition temperature $T_V$ at which the vacuum energy density
$\Lambda$ is induced, and with $G_{eff}$ being negative, we see that Eq.
(\ref{8a}) then only admits of solutions in which $k$ is negative, with the
global topology of the universe thus being fixed once and for all in the
conformal theory prior to the onset of any phase transition at all. Further,
with $G_{eff}$ being negative there is no initial singularity in the theory,
with the cosmology thus expanding from a finite minimum radius $R_{\rm min}$ and
a finite (though very large) maximum temperature $T_{\rm max}$. In such a
cosmology we find that at times which are not too early and not too late, the
solution to Eq. (\ref{8a}) can be well approximated as
\cite{Mannheim2000,Mannheim2001a}
\begin{equation}
R^2(t)=R_{\rm min}^2-kc^2t^2~~, 
\label{13a}
\end{equation}
corresponding to a curvature dominated universe with $k<0$. For such an
expansion radius we find that at recombination $d(\pi)$ and the horizon size
are well approximated as
\begin{equation}
d(\pi)=2t_R{\rm ln}\left({T_R \over T(t_0)}\right)~~,~~D(t_R,t=0)=t_R{\rm
ln}\left({2T_{\rm max}
\over T_R}\right)~~, 
\label{14a}
\end{equation}
with the theory thus being causally connected since $D(t_R,t=0)$ is
altogether bigger than $d(\pi)$. Thus in the conformal theory it will not be
fluctuations which start out at $t=0$ which will lead to anisotropies in the
CMB, but rather ones which start out at altogether later times, with explicit
calculation then showing that the fluctuations (of angular size
$\theta=(2T(t_0)/T_R){\rm ln} (T_F/T_R)$ when $\theta$ is small) which will
imprint a $1^{\circ}$ scale on the CMB would need to set out at a temperature of
order $10^{9}$~$^{\circ}$K, i.e. at just around the time of nucleosynthesis. In
the conformal theory then anisotropies in the CMB are still related to the size
of cosmological fluctuations but not to any horizon associated with them, with
it thus in principle being possible to produce the position of the first peak in
the CMB in theories in which $\Omega_k(t_0)$ is far from
negligible.\footnote{Determining whether the conformal theory can also fit the
shape and magnitude of the first peak or generate any secondary peaks has to
await the development of a full conformal fluctuation theory.}

The emergence of a nucleosynthesis scale for the onset of fluctuations is
not only intriguing,\footnote{Prior to the development of inflation it had
actually been thought that inhomogeneities would in fact first set in after (or
be catalyzed by) an epoch of homogeneous nucleosynthesis; and in principle
any such fluctuations could also be present in the standard theory
in addition to those generated by inflation itself.} it may also prove to be of
help for the conformal theory. Specifically, as noted earlier, one of the
outstanding challenges for conformal cosmology is that it while it can readily
synthesize the requisite amounts of primordial helium and lithium
\cite{Lohiya1999}, it fails to yield the needed amount of deuterium.
Specifically, because the cosmology expands at a much slower rate than the
standard theory, there is abundant time to destroy any deuterium which is
generated during nucleosynthesis, to thus leave the cosmology deuterium
deficient.\footnote{In compensation though it should be noted that because of
this very same slow expansion, and contrary to the situation in the standard
theory, conformal nucleosynthesis is able to not only get passed the $Z=8$ 
bottleneck of no stable nuclei with $Z=8$, but to then actually produce the
measured amount of $^9Be$ \cite{Lohiya1999}.} However, as noted by the authors
of \cite{Lohiya1999}, once helium is nucleosynthesized, the setting in of
inhomogeneities would lead to helium rich and helium deficient regions whose
spallation would then generate deuterium, with deuterium production then
occurring between nucleosynthesis and recombination. The emergence of
inhomogeneities toward the end of the nucleosynthesis era might then serve to
alleviate the conformal gravity deuterium problem. Development of a full
conformal cosmological fluctuation theory would allow one to address this
question while also making predictions for the CMB in a way which might
then prove definitive for the conformal theory.  The author is indebted to Dr.
R. A. Daly, Dr. K. Horne and M. P. Mory for their helpful comments and for
their active collaboration in this work. This work has been supported in  part
by the Department of Energy under grant No. DE-FG02-92ER40716.00.

\vfill\eject

\begin{table}
\begin{tabular}{llllll}
\hline
  & \tablehead{1}{r}{b}{54~SN~~~~~~~~}
  & \tablehead{1}{r}{b}{ 9~RG~~~~~~}
  & \tablehead{1}{r}{b}{54~SN~+~9~RG}
  & \tablehead{1}{r}{b}{20~RG~~~~~~}   
  & \tablehead{1}{r}{b}{54~SN~+~20~RG}\\
\hline
$\Omega_M$ & $0.30 \pm{}^{0.05}_{0.10}$ & $0.10 \pm{}^{0.40}_{0.20}$ & $0.25
\pm{}^{0.10}_{0.05}$
& $0.05 \pm{}^{0.25}_{0.05}$ & $0.25 \pm{}^{0.10}_{0.05}$ \\
M & $23.95 \pm 0.03$ &  & $23.93 \pm 0.03$
&  & $23.93 \pm 0.03$ \\
$\kappa$&  & $8.93 \pm 0.07$ & $8.82 \pm 0.07$ & $8.97 \pm 0.05$ &
$8.83 \pm 0.05$ \\
$\beta$ &  & $1.75 \pm 0.10$ & $1.8 \pm 0.10$ & $1.65 \pm 0.05$ &
$1.85 \pm 0.04$ \\
$\chi^2$ & 56.76 & 11.65 & 68.72 & 16.89 & 74.41 \\
DOF & 51 & 6 & 58 & 17 & 69 \\
\hline
\end{tabular}
\caption{Standard Cosmology with $k = 0$}
\label{tab1}
\end{table}

\begin{table}
\begin{tabular}{llllll}
\hline
  & \tablehead{1}{r}{b}{54~SN~~~~~~~~}
  & \tablehead{1}{r}{b}{ 9~RG~~~~~~}
  & \tablehead{1}{r}{b}{54~SN~+~9~RG}
  & \tablehead{1}{r}{b}{20~RG~~~~~~}   
  & \tablehead{1}{r}{b}{54~SN~+~20~RG}\\
\hline
$\Omega_M$ & $0.48 \pm{}^{0.10}_{0.38}$ & $0.00 \pm{}^{0.48}_{0.00}$ &
$0.38 \pm{}^{0.17}_{0.38}$ & $0.00 \pm{}^{0.24}_{0.00}$ &
$0.00 \pm{}^{0.45}_{0.00}$ \\
w & $-2.08 \pm{}^{1.39}_{0.92}$ & $-0.75 \pm{}^{0.43}_{1.28}$ &
$-1.36 \pm{}^{0.83}_{1.64}$ & $-0.73 \pm{}^{0.30}_{0.58}$ &
$-0.62 \pm{}^{0.09}_{1.18}$ \\
M & $23.91 \pm 0.03$ &   & $23.93 \pm 0.03$ &
 & $23.95 \pm 0.03$ \\
$\kappa$ &  & $8.88 \pm 0.07$ & $8.81 \pm 0.07$ & $8.88 \pm 0.05$ &
$8.81 \pm 0.05$ \\
$\beta$ &  & $ 1.75 \pm 0.10$ & $1.80 \pm 0.10$ & $1.70 \pm 0.04$ &
$1.75 \pm 0.04$ \\
$\chi^2$ & 56.18 & 11.43 & 68.56 & 16.53 & 74.09 \\
DOF & 50 & 5 & 57 & 16 & 68 \\
\hline
\end{tabular}
\caption{Quintessence with $k = 0$}
\label{tab2}
\end{table}

\begin{table}
\begin{tabular}{llllll}
\hline
  & \tablehead{1}{r}{b}{54~SN~~~~~~~~}
  & \tablehead{1}{r}{b}{ 9~RG~~~~~~}
  & \tablehead{1}{r}{b}{54~SN~+~9~RG}
  & \tablehead{1}{r}{b}{20~RG~~~~~~}   
  & \tablehead{1}{r}{b}{54~SN~+~20~RG}\\
\hline
$\Omega_M$
& $0.29 \pm{}^{0.08}_{0.24}$
& $0.05 \pm{}^{0.45}_{0.00}$
& $0.28 \pm{}^{0.08}_{0.23}$
& $0.05 \pm{}^{0.24}_{0.00}$
& $0.05 \pm{}^{0.29}_{0.00}$
\\
$\alpha$
& $0.00 \pm{}^{5.65}_{0.00}$
& $1.15 \pm{}^{6.85}_{1.15}$
& $0.00 \pm{}^{5.85}_{0.00}$
& $0.90 \pm{}^{7.10}_{0.90}$
& $3.35 \pm{}^{2.60}_{3.35}$
\\
M
& $23.94 \pm 0.03$
& 
& $23.94 \pm 0.03$
& 
& $23.95 \pm 0.03$
\\
$\kappa$
& 
& $8.89 \pm 0.07$
& $8.81 \pm 0.07$
& $8.90 \pm 0.05$
& $8.81 \pm 0.05$
\\
$\beta$
& 
& $1.75 \pm 0.10$
& $1.80 \pm 0.10$
& $1.70 \pm 0.04$
& $1.80 \pm 0.03$
\\
$\chi^2$
& 56.72
& 11.54
& 68.63
& 16.73
& 74.14
\\
DOF & 50 & 5 & 57 & 16 & 68 \\
\hline
\end{tabular}
\caption{Scalar Field Model with $k = 0$}
\label{tab3}
\end{table}

\begin{table}
\begin{tabular}{llllll}
\hline
  & \tablehead{1}{r}{b}{54~SN~~~~~~~~}
  & \tablehead{1}{r}{b}{ 9~RG~~~~~~}
  & \tablehead{1}{r}{b}{54~SN~+~9~RG}
  & \tablehead{1}{r}{b}{20~RG~~~~~~}   
  & \tablehead{1}{r}{b}{54~SN~+~20~RG}\\
\hline
$q_0$ & $-0.38 \pm{}^{0.38}_{0.17}$ & $-0.53 \pm{}^{0.53}_{0.47}$ &
$-0.38 \pm{}^{0.38}_{0.17}$
& $-0.50 \pm{}^{0.50}_{0.50}$ & $-0.38 \pm{}^{0.38}_{0.17}$ \\
M & $23.95 \pm 0.03$ &  & $23.95 \pm 0.03$
&  & $23.95 \pm 0.03$ \\
$\kappa$&  & $8.85 \pm 0.07$ & $8.81 \pm 0.07$ & $8.85 \pm 0.05$ &
$8.81 \pm 0.05$ \\
$\beta$ &  & $1.75 \pm 0.10$ & $1.8 \pm 0.10$ & $1.70 \pm 0.04$ &
$1.70 \pm 0.03$ \\
$\chi^2$ & 57.62 & 11.41 & 69.05 & 16.46 & 74.11 \\
DOF & 51 & 6 & 58 & 17 & 69 \\
\hline
\end{tabular}
\caption{Conformal Cosmology}
\label{tab4}
\end{table}

\begin{table}
\begin{tabular}{llllll}
\hline
  & \tablehead{1}{r}{b}{54~SN~~~~~~~~}
  & \tablehead{1}{r}{b}{ 8~RG~~~~~~}
  & \tablehead{1}{r}{b}{54~SN~+~8~RG}
  & \tablehead{1}{r}{b}{19~RG~~~~~~}   
  & \tablehead{1}{r}{b}{54~SN~+~19~RG}\\
\hline
$\Omega_M$ & $0.30 \pm{}^{0.05}_{0.10}$ & $0.15 \pm{}^{0.60}_{0.20}$ &
$0.30 \pm{}^{0.05}_{0.10}$ & $0.10 \pm{}^{0.25}_{0.10}$ &
$0.25 \pm{}^{0.10}_{0.05}$ \\
M & $23.95 \pm 0.03$ &  & $23.95 \pm 0.03$ &  & $23.93
\pm 0.03$ \\
$\kappa$ &  & $8.80 \pm 0.07$ & $8.70 \pm 0.07$ &
$8.86 \pm 0.05$ &
$8.75 \pm 0.05$ \\
$\beta$ &  & $1.55 \pm 0.11$ & $1.55 \pm 0.11$ & $1.60 \pm 0.05$ &
$1.70 \pm 0.03$ \\
$\chi^2$ & 56.76 & 2.20 & 59.06 &7.72 & 64.95 \\
DOF & 51 & 5 & 57 & 16 & 68 \\
\hline
\end{tabular}
\caption{Standard Cosmology with $k = 0$ and Outlier 3C 427.1 Removed}
\label{tab5}
\end{table}

\begin{table}
\begin{tabular}{llllll}
\hline
  & \tablehead{1}{r}{b}{54~SN~~~~~~~~}
  & \tablehead{1}{r}{b}{ 8~RG~~~~~~}
  & \tablehead{1}{r}{b}{54~SN~+~8~RG}
  & \tablehead{1}{r}{b}{19~RG~~~~~~}   
  & \tablehead{1}{r}{b}{54~SN~+~19~RG}\\
\hline
$\Omega_M$
& $0.48 \pm{}^{0.10}_{0.38}$
& $0.00 \pm{}^{0.76}_{0.00}$
& $0.45 \pm{}^{0.12}_{0.45}$
& $0.00 \pm{}^{0.31}_{0.00}$
& $0.02 \pm{}^{0.50}_{0.02}$
\\
w
& $-2.08 \pm{}^{1.39}_{0.92}$
& $-0.68 \pm{}^{0.53}_{2.32}$
& $-1.80 \pm{}^{1.23}_{1.20}$
& $-0.69 \pm{}^{0.32}_{0.76}$
& $-0.63 \pm{}^{0.11}_{2.06}$
\\
M
& $23.91 \pm 0.03$
& 
& $23.91 \pm 0.03$
& 
& $23.95 \pm 0.03$
\\
$\kappa$
& 
& $8.76 \pm 0.07$
& $8.72 \pm 0.07$
& $8.78 \pm 0.05$
& $8.74 \pm 0.05$
\\
$\beta$
& 
& $1.55 \pm 0.11$
& $1.55 \pm 0.10$
& $1.60 \pm 0.05$
& $1.65 \pm 0.05$
\\
$\chi^2$
& 56.18
& 2.14
& 58.66
& 7.32
& 64.79
\\
DOF & 50 & 4 & 56 & 15 & 67 \\
\hline
\end{tabular}
\caption{Quintessence with $k = 0$ and Outlier 3C 427.1 Removed}
\label{tab6}
\end{table}

\begin{table}
\begin{tabular}{llllll}
\hline
  & \tablehead{1}{r}{b}{54~SN~~~~~~~~}
  & \tablehead{1}{r}{b}{ 8~RG~~~~~~}
  & \tablehead{1}{r}{b}{54~SN~+~8~RG}
  & \tablehead{1}{r}{b}{19~RG~~~~~~}   
  & \tablehead{1}{r}{b}{54~SN~+~19~RG}\\
\hline
$\Omega_M$
& $0.29 \pm{}^{0.08}_{0.24}$
& $0.05 \pm{}^{0.69}_{0.00}$
& $0.28 \pm{}^{0.09}_{0.23}$
& $0.05 \pm{}^{0.30}_{0.00}$
& $0.05 \pm{}^{0.30}_{0.00}$
\\
$\alpha$
& $0.00 \pm{}^{5.65}_{0.00}$
& $1.90 \pm{}^{6.10}_{1.90}$
& $0.00 \pm{}^{5.65}_{0.00}$
& $1.35 \pm{}^{6.65}_{0.00}$
& $3.40 \pm{}^{2.65}_{3.40}$
\\
M
& $23.94 \pm 0.03$
& 
& $23.94 \pm 0.03$
& 
& $23.95 \pm 0.03$
\\
$\kappa$
& 
& $8.76 \pm 0.07$
& $8.71 \pm 0.07$
& $8.80 \pm 0.05$
& $8.74 \pm 0.05$
\\
$\beta$
& 
& $1.55 \pm 0.11$
& $1.55 \pm 0.11$
& $1.60 \pm 0.05$
& $1.65 \pm 0.04$
\\
$\chi^2$
& 56.72
& 2.16
& 58.99
& 7.51
& 64.78
\\
DOF & 50 & 4 & 56 & 15 & 67 \\
\hline
\end{tabular}
\caption{Scalar Field Model with $k = 0$ and Outlier 3C 427.1 Removed}
\label{tab7}
\end{table}

\begin{table}
\begin{tabular}{llllll}
\hline
  & \tablehead{1}{r}{b}{54~SN~~~~~~~~}
  & \tablehead{1}{r}{b}{ 8~RG~~~~~~}
  & \tablehead{1}{r}{b}{54~SN~+~8~RG}
  & \tablehead{1}{r}{b}{19~RG~~~~~~}   
  & \tablehead{1}{r}{b}{54~SN~+~19~RG}\\
\hline
$q_0$
& $-0.38 \pm{}^{0.38}_{0.17}$
& $-0.45 \pm{}^{0.45}_{0.55}$
& $-0.38 \pm{}^{0.38}_{0.17}$
& $-0.35 \pm{}^{0.35}_{0.65}$
& $-0.38 \pm{}^{0.38}_{0.17}$ \\
M
& $23.95 \pm 0.03$
& 
& $23.95 \pm 0.03$
& 
& $23.95 \pm 0.03$ \\
$\kappa$
& 
& $8.74 \pm 0.07$
& $8.72 \pm 0.07$
& $8.73 \pm 0.05$
& $8.74 \pm 0.05$ \\
$\beta$
& 
& $1.55 \pm 0.11$
& $1.55 \pm 0.11$
& $1.60 \pm 0.04$
& $1.60 \pm 0.03$ \\
$\chi^2$
& 57.62
& 2.13
& 59.76
& 7.20
& 64.83 \\
DOF & 51 & 5 & 57 & 16 & 68 \\
\hline
\end{tabular}
\caption{Conformal Cosmology with Outlier 3C 427.1 Removed}
\label{tab8}
\end{table}

\begin{figure}[!b]
\resizebox{.9\columnwidth}{!}
 {\includegraphics{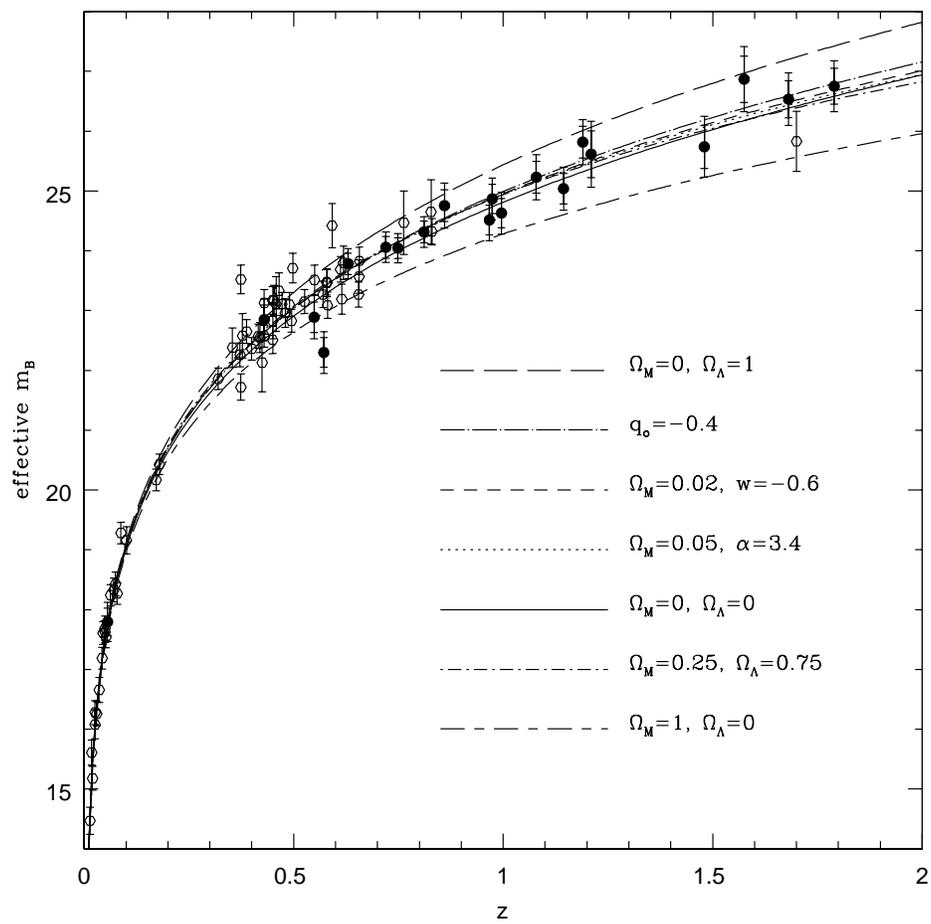}}
  \caption{Equivalent Hubble plot for supernovae (open circles)
and radio galaxies (closed circles).}
\label{fig1}
\end{figure}

\begin{figure}[!b]
\resizebox{.9\columnwidth}{!}
 {\includegraphics{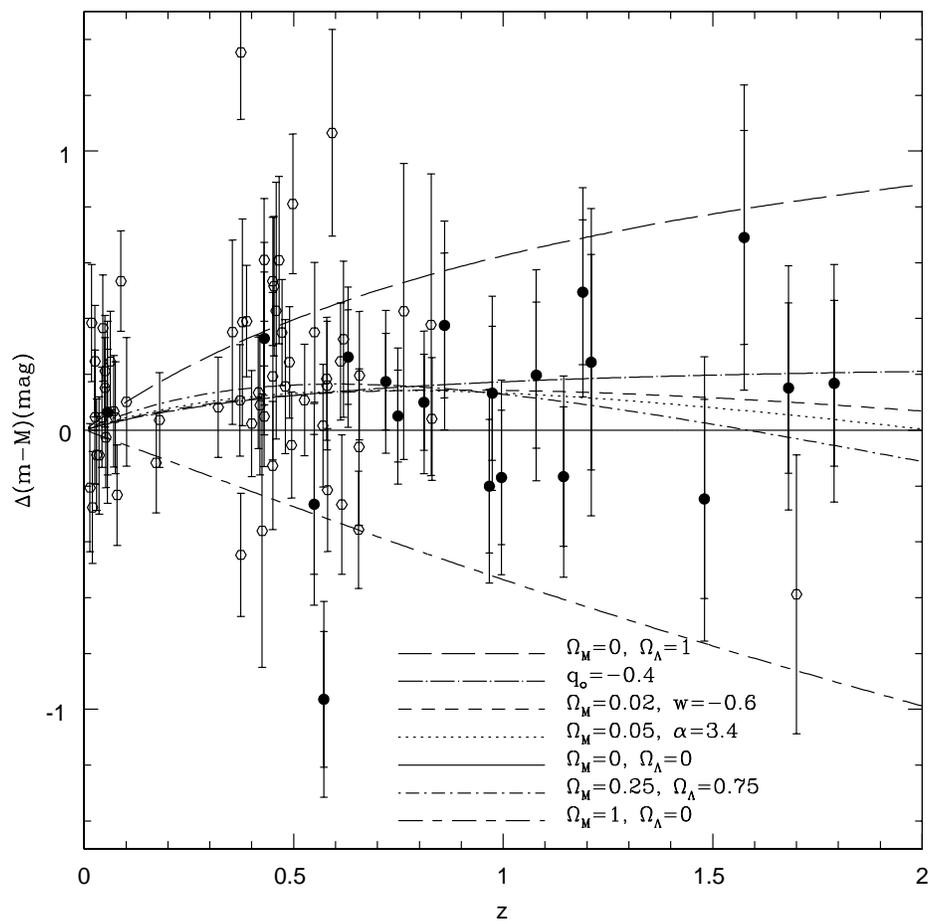}}
  \caption{Equivalent residual apparent magnitudes with respect to an empty
universe.}
\label{fig2}
\end{figure}

\end{document}